\documentclass[useAMS,usenatbib]{mn2e}
\usepackage{graphicx}                   
\usepackage{float}
\usepackage{amsmath}
\usepackage{amssymb}
\usepackage{multirow}
\usepackage{bm}
\usepackage{color}

\input{epsf.sty}                        
\input{psfig.sty}
\begin{document}

\title[Pulsar glitches in a strangeon star model. II. The activity]
{Pulsar glitches in a strangeon star model. II. The activity}

\author[Wang et al.]{W. H. Wang$^{1}$\thanks{E-mail: wang-wh@pku.edu.cn}, X. Y. Lai$^{2}$, E. P. Zhou$^{3}$, J. G. Lu$^{4}$, X. P. Zheng$^{5,6}$ and R. X. Xu$^{1,7}$\thanks{E-mail:r.x.xu@pku.edu.cn}\\
$^1$School of Physics, Peking University, Beijing 100871, China\\
$^2$Department of Physics and Astronomy, Hubei University of Education, Wuhan 430205, China\\
$^3$Max Planck Institute for Gravitational Physics (Albert Einstein Institute), Am M\"uhlenberg 1, Postdam-Golm 14476, Germany\\
$^4$CAS Key Laboratory of FAST, NAOC, Chinese Academy of Sciences, Beijing 100101, China\\
$^5$Institute of Astrophysics, Central China Normal University, Wuhan 430079, China\\
$^6$School of Physics, Huazhong University of Science and Technology, Wuhan 430079, China\\
$^7$Kavli Institute for Astronomy and Astrophysics at Peking University, Beijing 100871, China}

\maketitle

\begin{abstract}
Glitch is supposed to be a useful probe into pulsar's interior, but the underlying physics remains puzzling. The glitch activity may reflect a lower limit of the crustal moment of inertia in conventional neutron star models. Nevertheless, its statistical feature could also be reproduced
in the strangeon star model, which is focused here. We formulate the glitch activity of normal radio pulsars under the framework of starquake of solid strangeon star model, the shear modulus of strangeon matter is constrained to be $\mu\simeq 3\times10^{34}~\rm erg/cm^{3}$, consistent with previous work. Nevertheless, about ten times the shift in oblateness accumulated during glitch interval is needed to fulfill the statistical observations. The fact that typical glitch sizes of two rapidly evolving pulsars (the Crab pulsar and PSR B0540-69) are about two orders of magnitude lower than that of the Vela pulsar, significantly lower than the oblateness change they can supply, indicates probably that only a part of oblateness change is relieved when a pulsar is young. The unreleased oblateness and stress may relax as compensation in the following evolution. The small glitch sizes and low glitch activity of the Crab pulsar can be explained simultaneously in this phenomenological model. Finally, we obtain energy release to be $\Delta E\sim 2.4\times 10^{40}~\rm erg$ and $\Delta E\sim 4.2\times 10^{41}~\rm erg$ for typical glitch size of $\Delta\nu/\nu\sim 10^{-6}$ (Vela-like) and $\sim 10^{-8}$ (Crab-like). The upcoming SKA may test this model through the energy release and the power-law relation between the reduced recovery coefficient $Q/|\dot\nu|^{1/2}$ and $\Delta\nu/\nu$.
\end{abstract}

\begin{keywords}
dense matter-stars: neutron-pulsars: general.
\end{keywords}

\section{Introduction}
\label{Introduction}
Glitch is an abruptly spin-up phenomenon in rotation frequency of pulsars,
followed by a long timescale (tens of days to hundreds of days) relaxation towards the
pre-glitch state and accompanied by an increase in spin down rate in most cases.
Ever since its first discovery in the Vela pulsar~\citep{Radhakrishnan1969, Reichley1969}, more than 500
glitches have been reported in 190 pulsars~\footnote{http://www.jb.man.ac.uk/pulsar/glitches/gTable.html,
and http://www.atnf.csiro.au/people/pulsar/psrcat/glitchTbl.html.}.
The growing number of glitches has allowed statistical research on, e.g., glitch size distribution~\citep{Lyne2000, Wang2000, Yuan2010, Espinoza2011, Yu2013, Eya2019}, size-waiting time
correlation~\citep{Haskell2015, Ferdman2018, Antonopoulou2018, Melatos2018, Eya2019} and
glitch activity~\citep{McKenna1990, Lyne2000, Espinoza2011, Fuentes2017}.
These works help to understand the glitch phenomenon more comprehensively.

Currently, there are mainly two models developed to account for the glitch phenomenon, the starquake model and
the superfluid vortex model.
The starquake model was proposed soon after the discovery of the first glitch by Ruderman,
it attributes glitch to starquake in the solid crust of neutron stars when the stress builds
up during normal spin down reaches the critical value over which the star breaks down~\citep{Ruderman1969}.
But this model encounters difficulty in explaining large glitches in the Vela pulsar (glitch size $\Delta\nu/\nu \sim 10^{-6}$), as it predicts glitch recurrence time of human lifetime timescale~\citep{Baym1971}, which is obviously inconsistent with observations.
In the superfluid vortex model, glitch results from the sudden angular momentum transfer between
the faster-rotating superfluid interior and the crust (and that coupled to it) when the spin lag
reaches a critical value~\citep{Anderson1975,Alpar1984}, the year-long post-glitch
recovery timescale was interpreted as a strong evidence for superfluidity existence~\citep{Baym1969}, and the several recovery timescales indicates at least two-superfluid components in
the neutron star (NS) crust~\citep{Flanagan1990}.
It is worth mentioning that, starquake can act as the trigger for the superfluid vortex model~\citep{Akbal2018}, besides, the description of post-glitch features in some glitches needs the
combination of starquake and vortex model~\citep{Alpar1996, Akbal2015}.
At present, no conclusive evidence can rule out any of these two models.

Glitch has long been supposed to be a probe into pulsar's interior.
Given the absence of radiative and pulse profile changes in both radio and X-ray bands during
glitches (for the most recent observational results of the Crab pulsar, see ~\citep{Shaw2018, Zhang2018, Vivekanand2020}), and
the fact that glitches have never been observed in other celestial bodies, it is widely accepted
that glitches result from NS interior physics~\citep{Chamel2016}.
From this aspect, the vortex model has achieved great success in explaining post-glitch
recovery of Vela-like glitches.
Especially, glitch activity of single pulsar, defined following previous work~\citep{Link1999},
\begin{equation}
\label{eq1}
\langle A \rangle=\frac{\sum \Delta\nu/\nu}{t_{\rm{obs}}},
\end{equation}
has been considered as a parameter reflecting the lower limit of the crustal
superfluid reservoir $G$ (namely, fractional moment of inertia (MoI) of the crust) through
\begin{equation}
G=2\tau_{c}\langle A \rangle=\frac{\sum \Delta\nu}{t_{\rm{obs}}}/|\dot{\nu}|
\end{equation}
if the NS core does not contribute to the glitch, where $\nu$ is the pulsar spin frequency,
$\Delta\nu$ is the frequency increase during glitch,
$t_{\rm{obs}}$ is the accumulated observation time of pulsar in unit of years,
and $\tau_{c}=\nu/(2|\dot{\nu}|)$ is the pulsar characteristic age.
Link et al.~\citep{Link1999} found that the theoretical crustal moment of inertia matches well with the above MoI
requirement for $5$ frequently glitch pulsars (PSRs J1341-6220, J0835-4510, J1740-3015, 1826-1334 and J1801-2304), supporting the
superfluid vortex glitch model.
This result was further employed to set constraints on NS mass and its equation of state
(EoS)~\citep{Link1999, Ho2015}.

However, recent new observations are challenging the superfluid vortex model.
Firstly, glitches in magnetars and in the high magnetic field pulsars PSRs J1119-6127 and
J1846-0258 are occasionally accompanied by radiative changes~\citep{Akbal2015,Livingstone2010, Weltevrede2011, Archibald2016}, it is thus interpreted that
glitches in these pulsars could have a different physical origin~\citep{Dib2014, Kaspi2017}.
Secondly, single pulse observation of the Vela pulsar detected sudden changes in the pulse
shape coincident with the 2016 Vela pulsar glitch and was interpreted as alteration of the magnetosphere~\citep{Palfreyman2018}.
Most recently, Feng et al. declared strong association between soft X-ray polarization change
and the glitch of the Crab pulsar occurred on 23 July 2019~\citep{Feng2020}.
These most updated and growing observations show clearly that, glitch can induce radiative and/or
pulse profile changes in normal radio pulsars, high magnetic field pulsars and magnetars in X-ray
and/or radio bands.
These new observations probably indicates an unified physical origin of glitch.

In spite of the theoretical difficulty in explaining the above observations, for the superfluid vortex model,
whether the crust is enough or not is still under great debate when taking into account the non-dissipative
entrainment effect~\citep{Chamel2005, Carter2005a, Carter2005b, Carter2006, Chamel2006, Chamel2012} in NS crusts~\citep{Andersson2012, Chamel2013, Li2016, Wlazlowski2016, Watanabe2017, Basu2018}.

Whether the superfluid vortex model or starquake model, it is actually a matter of the nature of pulsars.
The neutron star model composed of neutron-rich matter is more popular at present, but the strangeon star
composed of solid quark-clusters could also exist based on phenomenological analysis and comparison with
observations~\citep{Xu2003, Lai2009, Lai2017}.

In this article, we explore how to describe the glitch activity in the solid strangeon star,
estimate the relevant physical parameters (the shear modulus) to fulfill the glitch activity statistical requirement and
try to improve the starquake model in strangeon stars.
Besides, much attention has been paid to the small typical glitch size and low glitch activity of the Crab-like young pulsars.

\section{A brief review on previous starquake model in strangeon stars}
The starquake model was first proposed by Ruderman under the framework of neutron star solid crust~\citep{Ruderman1969}.
The equilibrium configuration of a rotating incompressible fluid star is
generally described by the Maclaurin ellipsoid.
The ellipticity $e$ of a star with an average density $\overline\rho$  depends on its angular spin velocity $\Omega$ through~\citep{Chandrasekhar1969}
\begin{equation}
\label{mac-elli}
\Omega^{2}=2\pi G \overline\rho \bigg[ \frac{\sqrt{1-e^{2}}}{e^{3}}(3-2e^{2}) \sin^{-1}e-\frac{3(1-e^{2})}{e^{2}}\bigg],
\end{equation}
for slow rotators (i.e., ellipticity $e$ is small), Eq.(\ref{mac-elli}) is approximated to be
\begin{equation}
\Omega=2e\sqrt{\frac{2\pi G\overline\rho}{15}},
\end{equation}
where $G$ is the gravitational constant.
The moment of inertia of a non-rotating incompressible star, $I_{0}$, and the moment of
inertia of a rotating star, $I$, has the relation $I=I_{0}(1+\varepsilon)$.
$\varepsilon$ is the oblateness of the star, for slow ratators, $\varepsilon$ is defined as
\begin{equation}
\label{eq5}
\varepsilon=\frac{1}{3}e^{2}=\frac{5\Omega^{2}}{8\pi G \overline\rho}.
\end{equation}
As the star spins down, its oblateness and moment of inertia decrease, tending to readjust the stellar shape from oblate
toward spherical, but the rigidity of the solid crust resists this change, leaving
the crust remain more oblate than it would be if no resistance exists.
Stress develops during the resistance until it reaches the critical stress the crust can support.
The following sudden relaxation of stress will result in changes
in stellar shape and moment of inertia of the crust, namely, a glitch
occurs.
Baym \& Pines~\citep{Baym1971} developed this model and parameterized the dynamics in
NS solid crust.
According to this work, the total energy of the solid star is
\begin{equation}
\label{eqe}
E_{\rm{total}}=E_{0}+\frac{L^{2}}{2I}+A\varepsilon^{2}+B(\varepsilon-\varepsilon_{0})^{2},
\end{equation}
where $E_{0}$ is the total energy of a non-rotating star, $L^{2}/2I$ is the rotating energy,
$L$ is the total angular momentum, $I$ is the total moment of inertia, $A\varepsilon^{2}$ is modification
of gravitational energy of an ellipsoid relative to a spheroid star with the same mass and density,
\begin{equation}
\label{aeq}
A=\frac{3}{25}\frac{GM^{2}}{R}=6.21\times10^{52}(\frac{M}{1.4~M_\odot})^{2}(\frac{R}{10~\rm km})^{-1}~\rm erg,
\end{equation}
where $M$ is the stellar mass, $M_\odot=1.99\times10^{33}~\rm g$, $R$ is the
stellar radius and $1~\rm erg=1~\rm g~cm^{2}~s^{-2}$.
Besides, the fourth term in Eq.(\ref{eqe}) is the elastic energy, $B=\mu V/2$, $\mu$ is the shear modulus~\citep{Baym1971} and $V=4\pi R^{3}/3$ is the star's volume,
\begin{equation}
\label{beq}
B=\frac{\mu V}{2}=2.09\times10^{18}\mu~(\frac{R}{10~\rm km})^{3}~\rm erg,
\end{equation} 
$\varepsilon_{0}$ is the reference oblateness.
The equilibrium oblateness of a solid star is obtained by minimizing the total energy with respect to $\varepsilon$, thus
\begin{equation}
\label{eqe2}
\varepsilon=\frac{\pi^{2}\nu^{2}}{A+B}\frac{\partial I}{\partial \varepsilon}+\frac{B}{A+B}\varepsilon_{0}.
\end{equation}
Thus, Eq.(\ref{eqe2}) can be rewritten as
\begin{equation}
\label{ob-equilibrium}
\varepsilon=\frac{I_{0}\pi^{2}\nu^{2}}{A+B}+\frac{B}{A+B}\varepsilon_{0},
\end{equation}
the reference oblateness $\varepsilon_{0}$ is obtained by ignoring the strain energy,
\begin{equation}
\label{ll}
\varepsilon_{0}=I_{0}\pi^{2}\nu_{0}^{2}/A,
\end{equation}
$\nu_{0}$ is the initial spin frequency of the pulsar.
It is worth noting that, Rudeman assumed entire relaxation of stress after each quake, but Baym \& Pines
proposed that only part of the stress is relieved in the quake, and that effects of
plastic flow are comparatively small~\citep{Baym1971}.

\begin{figure}
\centering\resizebox{\hsize}{!}{\includegraphics{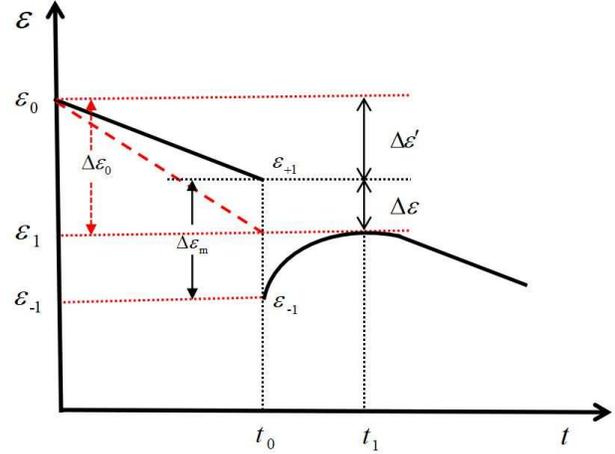}}
\caption{An illustration of the oblateness change as a function of time.
The black solid line represents oblateness of the solid star, while the red dashed line represents the oblateness of the Maclaurin ellipsoid.
}
\label{fig:oblateness}
\end{figure}

Based on quake model proposed by Ruderman and developed by Baym \& Pines, Zhou et al. developed
a starquake mechanism for pulsar glitches in the strangeon star model~\citep{Zhou2004}.
In Fig. \ref{fig:oblateness}, $\varepsilon_{0}$ is the initial reference oblateness, $\varepsilon_{+1}$/$\varepsilon_{-1}$ is the oblateness right before/after the glitch, and
$\varepsilon_{1}$ is the oblateness when the post-glitch recovery completes and the pulsar returns back to its steady state, $\varepsilon_{1}$ is the new reference oblateness for the following glitch.
$\Delta\varepsilon_{0}=\varepsilon_{0}-\varepsilon_{1}$ is the shift in reference oblateness, $\Delta\varepsilon_{\rm m}=\varepsilon_{+1}-\varepsilon_{-1}$ is the maximum shift in oblateness instantly after the glitch, $\Delta\varepsilon=\varepsilon_{+1}-\varepsilon_{1}$ is the shift in oblateness when the star reaches the steady state, $\Delta\varepsilon^{'}$ is the accumulated oblateness change during time interval $t_{q}$ between two successive glitches (or the time interval from the birth of the pulsar to the epoch of its first glitch) for a solid quark star, it equals $\Delta\varepsilon^{'}=|\dot\varepsilon|t_{q}$, $\Delta\varepsilon_{0}=\Delta\varepsilon+\Delta\varepsilon^{'}$.
In the normal spin down phase, the strangeon star changes the oblateness at the rate
\begin{equation}
\label{soliddeps}
\dot\varepsilon=\frac{2\pi^{2}I_{0}\nu\dot\nu}{A+B},
\end{equation}
while the reference Maclaurin ellipsoid changes the oblateness at the rate
\begin{equation}
\dot\varepsilon_{0}=\frac{2\pi^{2}I_{0}\nu\dot\nu}{A},
\end{equation}
therefore, oblateness of the solid star is always higher than the reference one during the normal spin down phase, stress will accumulate.
As illustrated in Fig.\ref{fig:oblateness}, the strangeon star is assumed to reach the critical stress at time $t_{0}$ and recover completely at time $t_{1}$.
The strangeon star readjusts the stellar shape and reduces its oblateness abruptly from $\varepsilon_{+1}$ toward $\varepsilon_{1}$ in a non-equilibrium way, a glitch occurs.
Its minimum oblateness, $\varepsilon_{-1}$, may be
well below $\varepsilon_{1}$ due to over recovery.
The physics behind over recovery is presented in paragraph 3 of section 4.1 in this paper.
After this, the oblateness of the strangeon star gradually increases from $\varepsilon_{-1}$ to $\varepsilon_{1}$.
The process of oblateness decrease from $\varepsilon_{+1}$ to $\varepsilon_{-1}$ corresponds to the glitch rise, while the process of oblateness increase from $\varepsilon_{-1}$ to $\varepsilon_{1}$ represents the post-glitch recovery process.

Zhou et al. found that the glitch size and the time intervals could be reproduced if the
strangeon star has a shear modulus $\mu=10^{30-34}~{\rm{erg/cm^{3}}}$ and critical stress
$\sigma_{c}=10^{18\sim24}~{\rm{erg/cm^{3}}}$~\citep{Zhou2004}, roughly consistent with Xu's estimation
that $\mu\sim 10^{32}~{\rm{erg/cm^{3}}}$ if oscillation of strangeon star is responsible
for the kilohertz quasi-periodic oscillations~\citep{Xu2003}.
It should be noted that, the strangeon star is totally solid, the so-called `glitch crisis' will not exist.

Peng \& Xu proposed two kinds of starquakes in the strangeon star model:
bulk-invariable (type I) and bulk-variable ones (type II)~\citep{Peng2008}.
A type I glitch occurs when the accumulated elastic energy exceeds the critical value
the star can stand against, while a type II glitch occurs when the star shrinks
its volume abruptly and this could be accretion-induced.
Zhou et al. parameterized the glitch size and energy release for both type I and II glitches~\citep{Zhou2014},
according to this work, a type I glitch has a size
\begin{equation}
\label{zhou14}
\frac{\Delta\nu}{\nu}=\Delta\varepsilon=\varepsilon_{+1}-\varepsilon_{1}=\frac{B}{A}\Delta\varepsilon^{'}.
\end{equation}
The glitch size can be expressed in another equivalent form according to~\citep{Baym1971},
\begin{equation}
\label{eqsize2}
\Delta\varepsilon=\frac{B}{A+B}\Delta\varepsilon_{0},
\end{equation}
where $\Delta\varepsilon_{0}=|\dot\varepsilon_{0}|t_{q}$.
If $B\simeq A$,
\begin{equation}
\label{eqzhou14}
\frac{\Delta\nu}{\nu}=\Delta\varepsilon\simeq \Delta\varepsilon^{'}.
\end{equation}
Note that $\Delta\varepsilon$ in Eq. (\ref{zhou14}) and Eq. (\ref{eqsize2}) correspond to the steady state of the pulsar, the corresponding glitch size is actually smaller than its real value instantly after the glitch because $\Delta\varepsilon<\Delta\varepsilon_{\rm m}$, as shown in Fig. \ref{fig:oblateness}.
Besides, a type II glitch has the glitch size
\begin{equation}
\frac{\Delta\nu}{\nu}=-\frac{\Delta I}{I}=-\frac{2\Delta R}{R},
\end{equation}
$\Delta I$ and $\Delta R$ are changes in moment of inertia and radius
during the glitch separately.

Lai et al.~\citep{Lai2018} further developed the starquake model by dividing motion of matter inside the star during starquake into the so-called plastic flow and elastic flow~(see Fig.2 in their work), which helps to explain the diversity of recovery coefficient $Q$ of all pulsar glitches.
According to this work, the total moment of inertia of a pulsar before glitch is given by
\begin{equation}
I=I_{0}(1+\varepsilon)(1+\eta),
\end{equation}
where $\eta$ describes the degree of density uniformity, $\eta=0$ if the star is density uniform.
The elastic flow occurs in the inner layer, results in decrease in the oblateness of the star, i.e., $\varepsilon$ decreases.
The plastic flow moves tangentially in the outer layer, leading to redistribution of matter from the equatorial region to the polar region~\citep{Franco2000}, and breaking of the density uniformity~\citep{Lai2018}.
Effect of matter motion towards higher latitudes is represented by decreases in $\eta$.
Both the plastic and the elastic flows contribute to the decrease of moment of inertia through
\begin{equation}
\label{plastic and elastic}
\Delta I\simeq -I_{0}\Delta\varepsilon-I_{0}\Delta\eta.
\end{equation}
In this model, the change in oblateness, $\Delta\varepsilon (>0)$, corresponds to the elastic motion, while the change in density uniformity, $\Delta\eta (>0)$,  corresponds to the plastic motion.
Both the elastic motion and the plastic motion contributes to the glitch size,
\begin{equation}
\label{gsize}
\frac{\Delta\nu}{\nu}=-\frac{\Delta I}{I}=\Delta\varepsilon_{\rm m}+\Delta\eta,
\end{equation}
$\Delta\varepsilon_{\rm m}$ is the maximum shift in oblateness instantly after the glitch, as defined in the second paragraph of section 2.
The plastic flow can not recover while the elastic flow can,
the recovery coefficient $Q$ is defined as
\begin{equation}
\label{Qdefinition}
Q=\frac{\varepsilon_{1}-\varepsilon_{-1}}{\Delta\varepsilon_{\rm m}+\Delta\eta}=\frac{\Delta\varepsilon_{\rm m}-\Delta\varepsilon}{\Delta\varepsilon_{\rm m}+\Delta\eta}.
\end{equation}
If $Q\ll 1$, the plastic flow dominates over the elastic flow, while if $Q\sim 1$, the elastic flow dominates over the plastic flow.
It should be noted that, the plastic flow is assumed to lead to no significant release of strain energy.

In this paper, we redefine contribution of oblateness change as `$\varepsilon$ effect', while contribution of matter motion induced change in density uniformity as `$\eta$ effect'.
We believe that both matter motion during oblateness change and matter motion from the equatorial region to the polar region are actually plastic flow, besides, the notion of elastic flow is misleading since the decrease in oblateness will actually not recover completely, as shown in Fig.\ref{fig:oblateness}.
Note that, these definitions are slightly different from that of Lai et al., however, it has no effect on the following calculations.

\section{glitch activity of strangeon stars}
In this section, we parameterize the glitch activity of strangeon stars based on the above starquake model.
The average glitch activity in statistical works~\citep{Lyne2000, Espinoza2011, Fuentes2017} is generally defined as
\begin{equation}
\dot{\nu}_{\rm g}=\frac{\sum_{i}\sum_{j}\Delta\nu_{ij}}{\sum_{i}T_{i}}
\end{equation}
for each group of pulsars divided according to
the spin down rate ($\rm log|\dot{\nu}|$), $\Delta\nu_{ij}$ represents change
in frequency due to glitch $j$ in pulsar $i$, and $T_{i}$ is the total
observation time over which pulsar $i$ has been searched for glitches.
Note that this definition is slightly different from that in Eq.(\ref{eq1}).
We adopt this definition hereafter in this article. Previously, using data covering 48
glitches in a total of 18 pulsars (289 pulsars monitored), Lyne et al. obtained
$\dot{\nu}_{\rm g}=(0.017\pm0.002)|\dot{\nu}|$ for pulsars with
$\tau_{c}>10^{4}~{\rm{yrs}}$, while the young Crab pulsar and
oldest pulsars, like the millisecond pulsars, have a low
glitch activity obviously departures from the linear
relation~\citep{Lyne2000}. Espinoza et al. continued this work
using a larger sample of 315 glitches in 102 pulsars
(more than 700 pulsars monitored) and claimed compatible results~\citep{Espinoza2011} with Lyne et al..
Most recently, Fuentes et al. extended the study
by Espinoza et al., using a database contains 384 glitches in 141 pulsars
(903 pulsars monitored), they obtained a best fitting
$\dot{\nu}_{\rm g}=(0.010\pm 0.001)|\dot{\nu}|$, which is consistent with the behavior of
all rotation-powered pulsars and magnetars with $-14<\rm log|\dot{\nu}|<-10.5$~\citep{Fuentes2017}. Similar to the results of Lyne et al.,
glitch activity of the rapidly evolving pulsars,
PSR B0540-69 (J0540-6919) and the Crab pulsar, do not follow the linear tendency.

In the following calculations, we try to reproduce the glitch activity presented by Fuentes et al.~\citep{Fuentes2017} for every single pulsar with $-14<\rm log|\dot{\nu}|<-10.5$ and explore the constraints on relevant physical inputs.
Pulsars with $\tau_{c}>10^{6}~\rm yrs$ are excluded
as their glitch activity departure from the linear tendency dramatically.
We do not consider error-bar in this fitting, i.e., we try to
reproduce the result $\dot{\nu}_{\rm g}=0.01|\dot{\nu}|$,
even though there exists individual calculation for 9 pulsars whose cumulative
fractional change of spin frequencies increase steadily with time~\citep{Ho2015}.
Besides, as stated above, type II glitches do not originate from the
oblateness change and may be induced by external accretion, we focus on
glitch activity of type I glitches in this paper and leave glitch activity of type II glitches for future works.

According to Lai et al.~\citep{Lai2018}, $\Delta\varepsilon_{\rm m}$ is the maximum oblateness change instantly after the glitch, which corresponds to the non-equilibrium state of the star, while $\Delta\varepsilon$ is the oblateness change when the recovery completes and the star returns to steady state.
Baym \& Pines~\citep{Baym1971} and Zhou et al.~\citep{Zhou2004} have shown that,
\begin{equation}
\label{bpe}
\Delta\varepsilon=\frac{B}{A+B}\Delta\varepsilon_{0}=\frac{B}{A+B}|\dot\varepsilon_{0}|t_{q}=\frac{B}{A(A+B)}2\pi^{2}I_{0}\nu|\dot\nu|t_{q}.
\end{equation}
Obviously, $\Delta\varepsilon_{\rm m}>\Delta\varepsilon$, we set
\begin{equation}
\Delta\varepsilon_{\rm m}=k\Delta\varepsilon, k>1,
\end{equation}
$k$ is the over recovery factor.
According to Eq. (\ref{gsize}) and Eq. (\ref{Qdefinition}), we arrive at
\begin{eqnarray}
&&\Delta\varepsilon_{\rm m}=\frac{kQ}{k-1}\frac{\Delta\nu}{\nu},
\nonumber\\
&&\Delta\varepsilon=\frac{Q}{k-1}\frac{\Delta\nu}{\nu},
\nonumber\\
&&\Delta\varepsilon_{\rm m}+\Delta\eta=\Delta\nu/\nu=\frac{(k-1)\Delta\varepsilon}{Q}.
\end{eqnarray}
The glitch activity for pulsar $i$ is
\begin{equation}
\dot{\nu}_{\rm g}=\frac{\nu \sum (\Delta\nu/\nu)}{T_{i}}=\frac{\nu}{T_{i}}(\sum \frac{(k-1)\Delta\varepsilon}{Q}),
\end{equation}
the summation runs over every glitches of pulsar $i$.
Since the recovery coefficient $Q$ for every glitches of pulsar $i$ is different, if we
assume the accumulative fractional change of every pulsar's spin frequency increases steadily, we will arrive at
\begin{equation}
\label{eqvgdot}
\dot{\nu}_{\rm g}=\frac{\nu}{T_{i}}(\sum \frac{(k-1)\Delta\varepsilon}{Q})=
\frac{\nu}{t_{q}}\frac{(k-1)\Delta\varepsilon}{Q}.
\end{equation}
Substitute Eq.(\ref{bpe}) into Eq.(\ref{eqvgdot}), the glitch activity in strangeon stars can be expressed as
\begin{equation}
\dot{\nu}_{\rm g}=(\frac{k-1}{Q}\frac{B}{A(A+B)}2\pi^{2}\nu^{2}I_{0})|\dot{\nu}|.
\end{equation}
For every glitch with measured recovery coefficient $Q$, parameter $k$
can be obtained using $\dot{\nu}_{\rm g}=0.01|\dot{\nu}|$,
if $A$ and $B$ are given. In the following calculations, we assume all pulsars
have the same mass $M=1.4~M_\odot$ and radius $R=10~\rm km$, thus
\begin{equation}
I_{0}=\frac{2}{5}MR^{2}=1.11\times10^{45}(\frac{M}{1.4~M_\odot})
(\frac{R}{10~\rm km})^{2}~\rm g~cm^{2}.
\end{equation}
Considering the recently reported heaviest millisecond pulsar J0740+6620 with mass
$2.14^{+0.10}_{-0.09}M_\odot$ ($68.3\%$ credibility interval)~\citep{Cromartie2019}, this simplicity amounts to
a largest uncertainty of the factor $0.52$.

We shall first discuss the relation between parameters $A$ and $B$.
In the neutron star model, only the thin outer crust is solid, therefore, it is generally believed that $B\ll A$.
However, for the strangeon star, the whole star is solid and its density is much higher than that of the solid crust of a neutron star, we thus expect higher shear modulus $\mu$ and parameter $B$ than that of the neutron star. Zhou et al.~\citep{Zhou2004} have found that $B/A\sim (10^{-4}-1)$ if the glitch size and time intervals are attibutes to starquake of the strangeon stars. In our calculations, if $A\gg B$, in order to reproduce the relation $\dot\nu_{\rm g}=0.01|\dot\nu|$,
\begin{equation}
k=1+\frac{0.01QA(A+B)}{B}\frac{1}{2\pi^{2}\nu^{2}I_{0}}\gg 1,
\end{equation}
the needed maximum oblateness change $\Delta\varepsilon_{\rm m}$ will be much larger than the oblateness change $\Delta\varepsilon$ in steady state, which is unrealistic.
We conjecture that, in order to reproduce the glitch activity $\dot\nu_{\rm g}=0.01|\dot\nu|$ in strangeon stars, parameter $B$ should at least be comparable with $A$, consistent with the discussion presented by Zhou et al.~\citep{Zhou2014}.
For a strangeon star, if $B\simeq A$, the shear modulus should be
\begin{equation}
\label{eq31}
\mu\simeq 3\times 10^{34}(M/1.4~M_\odot)^{2}(R/10~\rm km)^{-4}~\rm erg/cm^{3}.
\end{equation}
Note that, it is unlikely that parameter $B$ exceeds $A$, as a massive pulsar should be gravity bound.
Besides, the actual value of shear modulus of the strangeon star depends on properties of strangeon matter, which is highly uncertain and beyond the scope of this paper.

The high elastic energy comes from the large shear modulus of solid strange quark matter.
The state of matter inside pulsar-like compact stars depends on the challenging quantum chromodynamics (QCD), which is currently impossible to determine properties of QCD phase from the first principle.
Besides, calculations on solid state of quark matter at low temperature is much more difficult than that in liquid one.
In the crust of conventional NS, the shear modulus originated from coulomb interaction between lattice is~\citep{Clark1958}
\begin{equation}
\mu\sim (Ze)^{2}(m_{z})^{4/3},
\end{equation}
where $Ze$ is the electric charge of nuclei and $m_{z}$ is the nuclei density.
Similarly, the shear modulus originated only from electric interaction between charged n-quark clusters and an uniform background of electrons, and the shear modulus is well fitted by~(Strohmayer et al. 1991)
\begin{equation}
\label{eq33}
\mu\sim 0.12N(Z'e)^{2}/a\propto N^{4/3}(Z'e)^{2},
\end{equation}
where $Z'$ is the charge of quark-cluster, $N$ is the cluster's number density and $a$ is the separation between two nearby clusters.
It is apparent that $\mu$ is proportional to cluster's number density and the average charge of quark-cluster.
Note that, Eq.(\ref{eq33}) represents only the electric interaction, however, strong interaction dominates over coulomb interaction by several orders of magnitude, we thus expect that the smaller the dimensions of the cluster is, the larger the cluster's number density and the total shear modulus will be.
According to Zhou et al. 2004 (see Eq.(12) in their paper),
the lower limit of shear modulus of solid strange quark matter is $10^{28}~\rm{erg/cm^{3}}$, and the van der waals type color interaction with a high coupling constant may result in a much larger shear modulus and elastic energy.
Our estimations are consistent with the upper limit presented by Zhou et al., and our calculations are based on the conjecture $B\simeq A$.

Tables I and II show glitches with measured recovery coefficient\footnote{Data taken from the website http://www.atnf.csiro.au/people/pulsar/psrcat/glitchTbl.html.}
$Q\leq 1$ for pulsars with spin down rates $-14<\rm \log|\dot{\nu}|<-10.5$. The youngest glitching pulsars (PSR B0540-69 and the Crab pulsar)
and the old ones (PSRs J0528+2200, J1141-6545, J1812-1718 and J1853+0545) are not included.
PSR J1119-6127 is also excluded because it exhibits features of type II glitch.

\begin{table*} \centering \caption{Glitches with measured recovery coefficient $Q$}
\renewcommand{\arraystretch}{1.7}
\begin{tabular}{llllllllllllll}
\hline\hline
&$\rm{PSR}$         &$\rm{Glitch~No.}$       &$\rm{MJD}~(\rm{d})$     &$\Delta\nu/\nu~(10^{-9})$   &$\rm{\nu}~(\rm{Hz})$    &$\rm{|\dot{\nu}|}~(\rm{Hz/s})$   &$\tau_{c}~(\rm{kyr})$
&$Q$    &$k$\\
\hline
&J0205+6449     &1     &52920(144)   &5400(1800)    &15.2184     &4.48723E-11  &5.38      &0.77(11)             &187.18\\
\hline
&J0358+5413     &1     &46470(18)    &4366(1)       &6.39468     &1.79716E-13  &564.15    &0.00117(4)           &2.60\\
\hline
&J0631+1036     &1     &52852.50(1)  &17.6(1)       &3.47464     &1.26369E-12  &43.59     &0.62(5)              &2876.81\\
&               &2     &54632.530(2) &43.2(1)       &            &             &          &0.13(2)              &603.99\\
\hline
&J0835-4510     &1     &40280(4)     &2338(9)       &11.1982     &1.5675E-11   &11.33     &0.001980(18)         &1.88\\
&               &2     &41192(8)     &2047(30)      &            &             &          &0.00158(2)           &1.70\\
&               &3     &41312(4)     &12(2)         &            &             &          &0.1612(15)           &72.98\\
&               &4     &42683(3)     &1987(8)       &            &             &          &0.000435(5)          &1.19\\
&               &5     &43693(12)    &3063(65)      &            &             &          &0.00242(2)           &2.08\\
&               &6     &44888.4(4)   &1138(9)       &            &             &          &0.000813(8)          &1.36\\
&               &7     &45192.1(5)   &2051(3)       &            &             &          &0.002483(7)          &2.11\\
&               &8     &46259(2)     &1598.5(15)    &            &             &          &0.0037(5)            &2.65\\
&               &9     &47519.80360(8) &1805.2(8)   &            &             &          &0.005385(10)         &3.40\\
&               &10    &50369.345(2) &2110(17)      &            &             &          &0.030(4)             &14.39\\
&               &11    &51559.3190(5)&3152(2)       &            &             &          &0.0088(6)            &4.93\\
&               &12    &53193.09     &2100          &            &             &          &0.009(3)             &5.02\\
&               &13    &53959.93     &2620          &            &             &          &0.0119(6)            &6.31\\
&               &14    &57734.4855(4)&1433.87(2)    &            &             &          &0.0048(4)            &3.14\\
\hline
&J1048-5832     &1     &49034(9)     &2995(7)       &8.08407     &6.28179E-12  &20.40     &0.026(6)             &23.27\\
&               &2     &50788(3)     &771(2)        &            &             &          &0.008(3)             &7.85\\
\hline
&J1052-5954     &1     &54495(10)    &495(3)        &5.5371      &6.12668E-13  &143.29    &0.067(4)             &123.38\\
\hline
&J1112-6103     &1     &53337(30)    &1202(20)      &15.4083     &7.46674E-12  &32.72     &0.022(2)             &6.19\\
\hline
&J1123-6259     &1     &49705.87(1)  &749.12(12)    &3.6846      &7.13745E-14  &818.49    &0.0026(1)            &11.72\\
\hline
&J1301-6305     &1     &51923(23)    &4630(2)       &5.42005     &7.83602E-12  &10.97     &0.0049(3)            &10.34\\
\hline
&J1302-6350     &1     &50708.0(5)   &2.3(3)        &20.9205     &9.97313E-13  &332.59    &0.36(8)              &47.06\\
\hline
&J1341-6220     &1     &48645(10)    &990(3)        &5.17331     &6.77401E-12  &12.11     &0.016(2)             &34.48\\
&               &2     &50683(13)    &703(4)        &            &             &          &0.0112(19)           &24.43\\
\hline
&J1412-6145     &1     &51868(10)    &7253.0(7)     &3.17259     &9.96799E-13  &50.46     &0.00263(8)           &15.63\\
\hline
&J1420-6048     &1     &52754(16)    &2019(10)      &14.6628     &1.78806E-11  &13.00     &0.008(4)             &3.08\\
\hline
&J1531-5610     &1     &51731(51)    &2637(2)       &11.8765     &1.94566E-12  &96.78     &0.007(3)             &3.77\\
\hline
&J1702-4310     &1     &53943(169)   &4810(27)      &4.158       &3.86893E-12  &17.04     &0.023(6)             &75.49\\
\hline\hline
\end{tabular}\label{table1}
\end{table*}

\begin{table*}
\begin{center}\caption{Glitches with measured recovery coefficient $Q$, continued}
\renewcommand{\arraystretch}{1.7}
\begin{tabular}{llllllllllllll}
\hline\hline
&$\rm{PSR}$         &$\rm{Glitch~No.}$       &$\rm{MJD}~(\rm{d})$     &$\Delta\nu/\nu~(10^{-9})$   &$\rm{\nu}~(\rm{Hz})$    &$\rm{|\dot{\nu}|}~(\rm{Hz/s})$   &$\tau_{c}~(\rm{kyr})$
&$Q$    &$k$\\
\hline
&J1709-4429     &1     &48775(15)    &2057(2)       &9.76563     &8.86765E-12  &17.46     &0.01748(8)           &11.26\\
&               &2     &52716(57)    &2872(7)       &            &             &          &0.0129(12)           &8.57\\
&               &3     &54711(22)    &2743.9(4)     &            &             &          &0.00849(7)           &5.98\\
\hline
&J1730-3350     &1     &48000(10)    &3033(8)       &7.16846     &4.35909E-12  &26.07     &0.0077(5)            &9.39\\
&               &2     &52107(19)    &3202(1)       &            &             &          &0.0102(9)            &12.11\\
\hline
&J1731-4744     &1     &52472.70(2)  &126.4(3)      &1.20511     &2.37638E-13  &80.40     &0.073(7)             &2815.86\\
&               &2     &55735.18(14) &53.6(12)      &            &             &          &0.125(14)            &4820.97\\
&               &3     &56239.86(77) &10.7(17)      &            &             &          &0.14(10)             &5399.37\\
\hline
&J1740-3015     &1     &50941.6182(2)&1443.0(3)     &1.64772     &1.26553E-12  &20.64     &0.0016(5)            &34.00\\
&               &2     &52347.66(6)  &152(2)        &            &             &          &0.103(9)             &2125.51\\
&               &3     &53023.52     &1850.9(3)     &            &             &          &0.0302(6)            &623.91\\
&               &4     &58232.4(4)   &838.7(5)      &            &             &          &0.0068(4)            &141.26\\
\hline
&J1757-2421     &1     &55702(6)     &7815(3)       &4.27168     &2.37305E-13  &285.40     &0.0013(7)           &4.99\\
\hline
&J1801-2451     &1     &49476(3)     &1987.9(3)     &8.00641     &8.19935E-12  &15.48     &0.0050(19)           &5.37\\
&               &2     &52055(7)     &3755.8(4)     &            &             &          &0.024(5)             &21.97\\
&               &3     &54661(2)     &3101(1)       &            &             &          &0.0064(9)            &6.59\\
\hline
&J1803-2137     &1     &48245(11)    &4074.4(3)     &7.47943     &7.51635E-12  &15.78     &0.0137(3)            &14.71\\
&               &2     &50777(4)     &3184(1)       &            &             &          &0.0094(11)           &10.41\\
&               &3     &53429(1)     &3929.3(4)     &            &             &          &0.00630(16)          &7.31\\
\hline
&J1801-2304     &1     &53306.98(1)  &497(1)        &2.405      &6.53191E-13   &58.38     &0.009(2)             &88.14\\
\hline
&J1809-1917     &1     &53251(2)     &1625.1(3)     &12.0919    &3.73284E-12   &51.36     &0.00602(9)           &3.31\\
\hline
&J1826-1334     &1     &53737(1)     &3581(1)       &9.85222    &7.30442E-12   &21.39     &0.0066(3)            &4.81\\
\hline
&J1833-0827     &1     &48051(4)     &1865.6(1)     &11.7233    &1.26125E-12   &147.37    &0.0009(2)            &1.37\\
\hline
&J1841-0425     &1     &53388(10)    &578.6(3)      &5.37346    &1.84589E-13   &461.54    &0.00014(20)          &1.27\\
\hline
&J1844-0346     &1     &56135(7)     &3450(11)      &8.86525    &1.21583E-11   &11.56     &0.0145(22)           &11.33\\
\hline
&J1906+0722     &1     &55063(6)     &4538(14)      &8.96861    &2.88596E-12   &49.27     &0.0089(2)            &7.19\\
\hline
&J2337+6151     &1     &53615(6)     &20579.4(12)   &2.01857    &7.88237E-13   &40.60     &0.0046(7)            &64.22\\
\hline\hline
\end{tabular}
\end{center}
\label{table1}
{\sc Notes:} The first column shows pulsars' names, the second shows glitch numbers
for specific pulsars
with measured recovery coefficients, the third shows the glitch epochs,
the fourth shows the glitch sizes,  the fifth and sixth show the spin frequencies and
frequency derivatives, the seventh shows the characteristic ages, and the eighth and ninth show values of the recovery
coefficient $Q$ and the calculated parameter $k$. The number in bracket represent errorbar of the last significant digit.\\
\end{table*}

\begin{figure}
\centering
\includegraphics[width=0.53\textwidth]{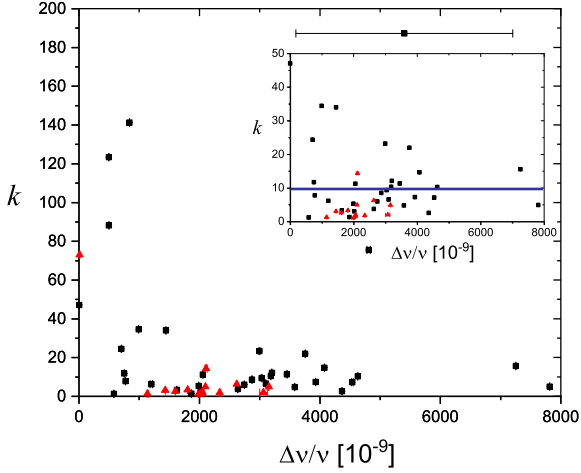}
\caption{The calculated parameter $k$ versus glitch size distribution.
The red triangles correspond to glitches  in the Vela pulsar, while the black squares correspond to glitches in all other pulsars in Tables I and II.
The inset shows where $k$ clusters, the blue horizontal line in the inset represents $k=10$.}
\label{fig:k}
\end{figure}

Fig. \ref{fig:k} shows distribution of $k$ values versus glitch sizes.
Clearly, $k$ values cluster around $k\sim 10$ for most of glitches, and have nearly no dependence on glitch sizes.
Results for the Vela pulsar may be pretty convincing, as most of its glitches have measured recovery coefficient, and its cumulative fractional change of spin frequency increases steadily with time.
Therefore we expect a nearly constant $k$ for all glitches in the Vela pulsar.
Indeed, $k\sim 5$ for glitches in the Vela pulsar except
the glitch occurred on MJD 41312, which means that $\Delta\varepsilon_{\rm m}\sim 5\Delta\varepsilon$ to explain glitch activity of a solid quark star with uniform density.
Considering its real glitch activity $\dot\nu_{\rm g}=(1.62\pm 0.03)\%|\dot\nu|$~\citep{Ho2015}, $k\sim 8$.
Besides, other pulsars whose cumulative fractional change in spin frequency increase steadily with time, e.g., PSRs J1709-4429, J1801-2451 and J1803-2137 also have $k$ clusters around $k\sim 10$. For several glitches, e.g., glitch in PSRs J0631+1036, J1731-4744 and J1740-3015, $k$ is systematically large and seems to be unrealistic, however, given the uncertainties in stellar mass, radius and especially possible systematical overestimation of glitch activity of individual pulsar (see the errorbars in Fig. 4 in the work by~\citealt{Fuentes2017}), these results could be understood.
To summarize, a consistent maximun oblateness change $\Delta\varepsilon_{\rm m}\sim 10\Delta\varepsilon$ is needed for most of glitches in radio pulsars.

\subsection{The puzzling glitch activity of the Vela pulsar}

As stated above, for the Vela pulsar,  $\Delta\varepsilon_{\rm m}\sim 5\Delta\varepsilon$ ($\Delta\varepsilon_{\rm m}\sim 8\Delta\varepsilon$) is needed to explain its $\dot\nu_{\rm g}=0.01|\dot\nu|$ ($\dot\nu_{\rm g}=(1.62\pm 0.03)\%|\dot\nu|$).
This result is puzzling, as more oblateness change than that accumulated during glitch interval is needed.
The extra oblateness change may simply come from over recovery when the oblateness decreases abruptly from $\varepsilon_{+1}$ to $\varepsilon_{-1}$, however, this explanation is hard to be tested as the over recovery corresponds to a non-equilibrium process and $\varepsilon_{-1}$ can not be analytically calculated principally.
What's more, even if over recovery is not considered, we think $\varepsilon$ effect may be enough to account for the glitch activity for the Vela pulsar, as estimated below.

Pulsars are supposed to spin very fast when they are first born, with an initial period possibly less than $1~\rm ms$~\citep{Lai2001}, which means that their initial oblateness is pretty large.
The following decrease in oblateness could support more than $10^{4}$ times large glitches with glitch size up to $10^{-6}$.
The glitch activity considering only $\varepsilon$ effect is
\begin{equation}
\label{lll}
\dot\nu_{\rm g}=\frac{\nu\sum (\Delta\nu/\nu)}{\Delta T}\geq \frac{2\tau_{c}}{\Delta T}(\frac{B}{A+B}\sum \Delta\varepsilon_{0})|\dot\nu|,
\end{equation}
where $\Delta T$ is the total observation time over which the pulsar has been searched for glitches, $\sum \Delta\varepsilon_{0}$ is the total oblateness change of a Maclaurin sphere over observation time $\Delta T$. Supposing the Vela pulsar has an initial spin frequency $\nu_{0}=1000~\rm Hz$, and a present spin frequency $\nu_{\rm vela}=11.1982~\rm Hz$.
According to Eq.(\ref{ll}),
\begin{equation}
\label{llgg}
\sum \Delta\varepsilon_{0}=\frac{\pi^{2}I_{0}}{A}(\nu_{0}^{2}-\nu_{\rm vela}^{2}).
\end{equation}
Extending the observation time to its birth, its long timescale glitch activity should be
\begin{equation}
\dot\nu_{\rm g-vela}\geq\frac{2\pi^{2}BI_{0}}{A(A+B)}(\nu_{0}^{2}-\nu_{\rm vela}^{2})|\dot\nu|.
\end{equation}
Using the typical NS mass $M=1.4~M_\odot$, $R=10~\rm km$ and the conjecture $B\simeq A$,
\begin{equation}
\dot\nu_{\rm g-vela}\geq 0.176|\dot\nu|\gg 0.01|\dot\nu|.
\end{equation}
Even if we set the initial spin frequency one order of magnitude lower~\citep{Xu2002}, the result is still close to $0.01|\dot\nu|$.
This above rough estimation demonstrates that, even if $\eta$ effect is neglected, glitch activity of the Vela pulsar
could at least be comparable to $0.01|\dot\nu|$ in timescale of its characteristic age.

We conclude here that, $\varepsilon$ effect may be sufficient to explain the glitch activity of the Vela pulsar in long timescale, but the oblateness change accumulated during glitch intervals are insufficient to explain the glitch activity statistics $\dot\nu_{\rm g}=0.01|\dot\nu|$ even when $B\simeq A$ for a solid quark star with uniform density under the framework of previously established starquake models using its real observation time $\Delta T\sim 50~\rm yr$, about $(5-8)$ times that accumulated during glitch intervals is needed.
But where can the extra oblateness change come from?
Before answering this question, let's take a look at the puzzling phenomena of the Crab pulsar.
The Crab and the Vela pulsars have similar glitch intervals, about $(2-3)~\rm yr$, but the Crab pulsar spins down a factor of $24$ faster than the Vela pulsar.
If the accumulated oblateness change during glitch intervals account for the glitch size, the glitch size and glitch activity of the Crab pulsar should be larger than that of the Vela pulsar.
What's puzzling is that, typical glitch size of the Crab pulsar is about two orders of magnitude lower than that of the Vela pulsar, besides, glitch activity of the Crab pulsar is also lower than that of the Vela pulsar.
It seems that only a small part of the accumulated oblateness change is relaxed during glitch in the Crab pulsar.
So where has the extra oblateness change gone?
These two questions may have some connections.
In the following subsection, we try to answer the small glitch size and low glitch activity of the Crab pulsar first under the framework of starquakes of solid strangeon star model.

\subsection{Why the glitch sizes of the Crab pulsar are small, and its glitch activity is low?}
Typical glitch size of the young Crab pulsar is about two orders of magnitude lower than that in the Vela pulsar,
and its glitch activity is $\dot\nu_{\rm g-crab}\sim 10^{-14}~\rm Hz/s\sim 2.7\times 10^{-5}|\dot\nu|_{\rm crab}$~\citep{Fuentes2017},
which is about three orders of magnitude lower than the linear trend $\dot\nu_{\rm g}=0.01|\dot\nu|$.
Similar features occur in another young pulsar PSR B0540-69, whose supernova remnant age equals to $1000^{+600}_{-240}~\rm yr$~\citep{Park2010}.

Our rough estimation shows that if $B\simeq A$, the shift in oblateness $\Delta\varepsilon$ during glitch interval is larger than glitch size of the Crab pulsar, and its glitch activity should be larger than what is observed from the point of view of starquake model of an uniform solid quark star.
According to Eq.(\ref{soliddeps}), if $B\simeq A$,
\begin{equation}
\dot\varepsilon \simeq \frac{\pi^{2}I_{0}\nu\dot\nu}{A}=1.76\times 10^{-7}\nu\dot\nu(\frac{M}{1.4~M_\odot})^{-1}(\frac{R}{10~\rm km})^{3}~\rm s^{2}.
\end{equation}
For the Crab pulsar, $\nu_{\rm crab}=29.9491~\rm Hz$, $\dot\nu_{\rm crab}=-3.776\times 10^{-10}~\rm Hz/s$ and its time interval since last glitch is roughly $\Delta t_{q-\rm crab}\sim 2~\rm yr$, $|\dot\varepsilon|_{\rm crab}=6.277\times 10^{-8}~\rm yr^{-1}$ if $M=1.4~M_\odot$, $R=10~\rm km$ and $B\simeq A$,
the absolute accumulated oblateness change during $\Delta t_{q-\rm crab}$ is
\begin{equation}
\Delta\varepsilon_{\rm crab}\sim |\dot\varepsilon|_{\rm crab}\Delta t_{q-\rm crab}=1.255\times10^{-7}.
\end{equation}
In this case, the glitch size should be $\Delta\nu/\nu=B\Delta\varepsilon_{\rm crab}/A\simeq \Delta\varepsilon_{\rm crab}\sim 1.255\times10^{-7}$ according to Eqs. (\ref{zhou14}) and (\ref{eqzhou14}), large enough to support a glitch of size $\Delta\nu/\nu\sim 10^{-8}$.
If only $\varepsilon$ effect is considered, for the Crab pulsar,
\begin{equation}
\label{v1v2}
\sum \Delta\varepsilon_{0}=\frac{\pi^{2}I_{0}}{A}(\nu_{1}^{2}-\nu_{2}^{2}),
\end{equation}
where $\nu_{1}$ and $\nu_{2}$ represent the initial and final spin frequencies of the Crab pulsar during the observation time $\Delta T$.
Rotation of the Crab pulsar has been monitored since 1986 by daily observations at Jodrell Bank Observatory using mainly the $13$-\rm m radio telescope at $610~\rm MHz$~\citep{Lyne1988, Lyne1993, Lyne2015}, its spin frequency on MJD 47926 is $\nu_{1}=29.9843723662~\rm Hz$ and $\nu_{2}=29.6122791665~\rm Hz$ on MJD 58917~\footnote{http://www.jb.man.ac.uk/research/pulsar/crab.html}.
Substituting $\nu_{1}$ and $\nu_{2}$ into Eq.(\ref{v1v2}), we get its total oblateness change during this period $\Delta T_{\rm crab}\simeq 32~\rm yr$,
\begin{equation}\
\label{sum-eps}
\sum \Delta\varepsilon_{0}= 3.9\times10^{-6}(\frac{M}{1.4~M_{\odot}})^{-1}(\frac{R}{10~\rm km})^{3},
\end{equation}
therefore, the lower limit of the glitch activity of the Crab pulsar should be
\begin{equation}
\label{crabngdot}
\dot\nu_{\rm g-crab-low}\geq \frac{2\tau_{c}B\sum \Delta\varepsilon_{0}}{\Delta T_{\rm crab}(A+B)}|\dot\nu|_{\rm crab}=1.54\times10^{-4}|\dot\nu|_{\rm crab}
\end{equation}
for a NS with $M=1.4~M_\odot$ and $R=10~\rm km$ under the assumption $B\simeq A$.
Even though $\eta$ effect is neglected, the lower limit is still a factor of $5.7$ larger than its statistical glitch activity $\dot\nu_{\rm g-crab}\sim 2.7\times 10^{-5}|\dot\nu|_{\rm crab}$.

So why glitch activity of the Crab pulsar is about three orders of magnitude lower than the linear trend, while the Vela pulsar falls on the linear trend perfectly? Comparison between the Vela and the Crab pulsars may uncover the secret.
The Vela pulsar has $\nu_{\rm vela}=11.1982~\rm Hz$, $\dot\nu_{\rm vela}=-1.5675\times 10^{-11}~\rm Hz/s$, $\Delta t_{q-\rm vela}\sim 3~\rm yr$,
$|\dot\varepsilon|_{\rm vela}\sim 2.93\times 10^{-9}~\rm yr^{-1}$ and $Q_{\rm vela}\sim (1-5)\times10^{-3}$, while the Crab pulsar
has $\nu_{\rm crab}=29.9491~\rm Hz$, $\dot\nu_{\rm crab}=-3.776\times10^{-10}~\rm Hz/s$, $|\dot\varepsilon|_{\rm crab}\sim 6.277\times 10^{-8}~\rm yr^{-1}$ and $Q_{\rm crab}\sim 1$~\footnote{https://www.atnf.csiro.au/people/pulsar/psrcat/glitchTbl.html}.
The biggest difference may be that, the Crab pulsar is spinning down so rapidly than the Vela pulsar that $\varepsilon$ effect is large enough to explain its glitch size and glitch activity, while the Vela pulsar needs $\eta$ effect.
It seems that only a small part of $\varepsilon$ effect contributes to the glitch size, and glitches occur before $\eta$ effect works in the Crab pulsar.

Inspired by this idea, we estimate $I_{\varepsilon}$, the fractional moment of inertia of $\varepsilon$ effect involved region that contributes to the glitch size in the Crab pulsar.
In order to reproduce the typical glitch size of $10^{-8}$,
\begin{equation}
\frac{\Delta\nu}{\nu}=-\frac{\Delta I}{I}\sim\frac{I_{\varepsilon}|\dot\varepsilon|_{\rm crab}\Delta t_{q-\rm crab}}{I}\sim 10^{-8}.
\end{equation}
We get $I_{\varepsilon}\sim 8\%I$.
On the other hand, to reproduce its glitch activity,
\begin{equation}
\label{pre-ts}
\dot\nu_{\rm g-crab}=\frac{I_{\varepsilon}}{I}\dot\nu_{\rm g-crab-low}\sim 2.7\times 10^{-5}|\dot\nu|_{\rm crab},
\end{equation}
we get
$I_{\varepsilon}\sim 17\%I$, in this case, typical glitch size of the Crab pulsar should be
\begin{equation}
\label{ts}
(\frac{\Delta\nu}{\nu})_{\rm typical}\sim 2\times 10^{-8}.
\end{equation}
The latter result, $I_{\varepsilon}\sim 17\%I$, may be more reliable, as the glitch activity averages over a long time span.
Note that, the estimated $I_{\varepsilon}$ has dependency on the mass and radius of the star.
By introducing the fractional moment of inertia of $\varepsilon$ effect involved region $I_{\varepsilon}\sim 17\%I$, we can simultaneously and phenomenologically answer why typical glitch size of the Crab pulsar is two orders of
magnitude lower that in the Vela pulsar, and why glitch activity of the Crab pulsar is about three orders of magnitude lower than
the linear relation $\dot\nu_{\rm g}=0.01|\dot\nu|$. 
Note that, $I_{\varepsilon}$ has mass and radius dependencies. According to Eqs.(\ref{sum-eps}), (\ref{crabngdot}) and (\ref{pre-ts}),
\begin{equation}
I_{\varepsilon}\sim 17\%(\frac{M}{1.4~M_\odot})(\frac{R}{10~\rm km})^{-3}.
\end{equation}

The next question is, why only a small fraction of the total moment of inertia contributes to the glitch size in the Crab pulsar?
The answer may lie in the density difference between the outer and inner layers of the star.
According to Eq. (\ref{soliddeps}), if $B\simeq A$, the change rate of oblateness is
\begin{equation}
\label{oblate-density}
|\dot\varepsilon|\sim\frac{\pi^{2}I_{0}\nu|\dot\nu|}{A}\propto \frac{I_{0}}{A}\propto 1/\overline\rho,
\end{equation}
which is inversely proportional to the density, thus the low density region changes the oblateness faster than the high density region.
The low density region in the Crab pulsar evolves faster thus it is easier to reach the critical stress than the high density region.
Given this, in the most youngest pulsars such as the Crab pulsar and PSR B0540-69, $\varepsilon$ effect in the low density region could account for its glitch size and glitch activity.
It is estimated that about $17\%$ of the total moment of inertia of the Crab pulsar contributes to its glitch size, while the slowly evolving inner part
which amounts to $83\%I$ is still accumulating stress and doesn't contribute to the small glitches.
However, in Vela-like middle-aged pulsar, the spin down rate has decrease dramatically, the difference in change rate of oblateness between the low and high density region is not as large as that in the Crab-like young pulsars, therefore, both the low density region and at least part of the high density region contribute to its glitch size.
This picture can reasonably explain where the extra oblateness change in the Crab pulsar has gone.
From the view of evolutionary, this extra oblateness may act as the source of the extra oblateness change the Vela pulsar needed.
Therefore, traditional starquake models should be modified.
An phenomenological two-layered starquake model is constructed in the following.

\section{Two-layered starquake model}
The rough two-layered starquake model of strangeon stars can be illustrated by Figs.\ref{fig:crab-like} and \ref{fig:vela-like}.
This model is established based on assumptions that $B\simeq A$ and there exists a density gradient in the star.

A Newly born pulsar is supposed to be liquid because of the high temperature (as high as $\sim 10^{11}~\rm K$) during supernova explosion, after the temperature has decreased below the melting temperature of strangeon matter as a result of cooling, the star solidifies and can be treated as a huge stone.
This cooling process before solidification may not last long, therefore, the outer low density layer has nearly the same initial oblateness, $\varepsilon_{0}$, with that of the inner high density layer, both of which are determined by the Maclaurin ellipsoid with an initial angular spin velocity $\Omega_{0}$ through Eq.(\ref{ll}).
$\varepsilon_{0}$ is the reference oblateness before the first glitch.

\subsection{starquake in Crab-like young pulsars}
\begin{figure}
\centering\resizebox{\hsize}{!}{\includegraphics{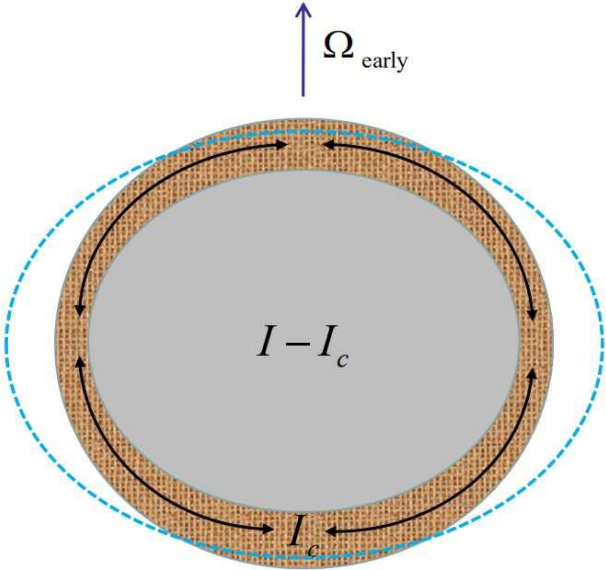}}
\caption{This figure shows the difference in oblateness between the small stones (khaki) in the outer part and the big stone (grey) in the inner part of a Crab-like young pulsar with the angular spin velocity $\Omega_{\rm early}$.
The outermost dashed ellipsoid (light blue) represents the initial reference oblateness $\varepsilon_{0}$.
The directions of the black solid arrows represent that, matter can move from the equatorial region to the polar region, if its kinetic energy is large enough, matter may move back towards the equatorial region after the collision in the polar region.
The upward blue arrow represents the direction of angular momentum.
Note that there is no need that $I_{\rm c}=I_{\varepsilon}$, but if the inner part contributes to $\varepsilon$ effect, $I_{\rm c}\geq I_{\varepsilon}$.}
\label{fig:crab-like}
\end{figure}

The change rate of oblateness is inversely proportional to density as shown in Eq. (\ref{oblate-density}), thus the low density layer
decreases the oblateness much faster than the high density layer.
As the star spins down, their difference in oblateness increases gradually.
After a period of time, the oblateness of the low density layer departures dramatically from the initial reference oblateness, while that of the high density layer remains close to the initial reference oblateness, stress accumulates mainly in the low density layer.
Once the critical stress is reached, the low density layer cracks first and results in a glitch, the detailed and complete process of a glitch is presented in the next paragraph.
This process breaks the low density layer of the huge stone into small pieces, or sand-like small stones, irreversibly, resulting in an increase in fluidity of solid matter.
As a consequence, the star can be divided into two parts structurally, the outer part composed of small stones, and the inner part which remains a big stone.
The moments of inertia of small stones and the remaining big stone are denoted by $I_{\rm c}$ and $(I-I_{\rm c})$ separately.
Note that, the big stone will also crack when the critical stress is reached, the outermost part of the big stone will break into small stones intermittently and contributes to the fractional change of moment of inertia, thus $I_{\rm c}$ increases with time in long timescale.
For Crab-like young pulsar with the angular spin velocity $\Omega_{\rm early}$ ($\Omega_{\rm early}\lesssim \Omega_{0}$), its structure is shown exaggeratedly in Fig.\ref{fig:crab-like} in order to make it apparent.

The detailed starquake process may be described hereafter.
The equatorial plane of the low density layer cracks when the stress reaches a critical value.
The overall effect is that, the star is compressed in the surface and becomes a bit more spherical, and the local density in the equatorial region increases slightly.
However, the cracking may reduce flatness of the surface and results in small-scale mountains.
As pointed out by Yim \& Jones, the process of mountain slowly dissipating away through plastic flow~\citep{Baiko2018} or magnetic diffusion~\citep{Pons2019} can explain the post-glitch recovery on a time-scale similar to the glitch recovery time-scale $\tau$~\citep{Yim2020}.

The compressed matter redistributes through two ways.
On the one hand, part of the matter around the cracking place is accelerated by the gravity and brought towards its new equilibrium oblateness $\varepsilon_{1}$ ($\varepsilon_{1}<\varepsilon_{0}$).
However, this process represents compression of the equatorial region, thus the mass density in the equatorial region increases slightly.
As shown in Eq.(\ref{eq5}), the equilibrium oblateness of a Maclaurin ellipsoid is inversely proportional to the average density, so the abrupt compression and density increase results in a decrease in the new equilibrium oblateness of the outer part then its original one, which explains the over recovery.
The effect of oblateness change is called $\varepsilon$ effect, as we defined in section 2, which represents the local density change in the equatorial region essentially.
Our estimations show that, if the density in the outer part of the Crab pulsar increases by a factor of $10^{-8}$, this increase will be enough to explain the over recovery factor $k\sim 10$.
On the other hand, another part of the matter around the cracking place may be pushed towards the polar region in a timescale possibily less than $12.6~\rm s$~\citep{Ashton2019}, the gravitational and the strain energy may provide its kinetic energy.
This process redistributes the matter and induces decrease in density uniformity, which contributes to the glitch size simultaneously.
The change in density uniformity is called $\eta$ effect, also defined in section 2.

This phenomenological two-layered starquake model can explain the frequent small glitches in the Crab pulsar.
As the low density layer cracks gradually, the small stones in the outer part will be loosely connected with each other and form a metastable structure, something like the seismic fault zone on the earth.
Once the critical stress is accumulated through oblateness decrease, the outer part cracks and a glitch will be triggered.
For Crab-like young pulsars, the metastable structure may overlap with the low density layer.
The oblateness of the metastable structure decreases quickly due to its large spin down rats, which may reproduce the frequent small glitches in the Crab pulsar.
As estimated in section 3.2, if only $\varepsilon$ effect is considered, a moment of inertia of about $17\%I$ is enough to explain the typical glitch size and the low glitch activity of the Crab pulsar simultaneously.

The inner big stone could also crack after a relatively long period of oblateness decrease than the outer part, and this could be used to explain the two relatively large glitches in the Crab pulsar.
The Crab pulsar has experienced  two relatively large glitches~\footnote{http://www.jb.man.ac.uk/pulsar/glitches/gTable.html} on MJD 53067.0780 ($\Delta\nu/\nu=2.14\times 10^{-7}$) and MJD 58064.555 ($\Delta\nu/\nu=5.16\times 10^{-7}$).
We assume the outer and inner parts crack when a constant oblateness change $\Delta\varepsilon_{\rm c}$ is reached.
Considering only $\varepsilon$ effect, the typical glitch size of the outer part is
\begin{equation}
(\frac{\Delta\nu}{\nu})_{\rm out}=\frac{I_{\rm c}\Delta\varepsilon_{\rm c}}{I}.
\end{equation}
Similarly, the typical glitch size of the inner part is
\begin{equation}
(\frac{\Delta\nu}{\nu})_{\rm in}=\frac{(I-I_{\rm c})\Delta\varepsilon_{\rm c}}{I}=\frac{I-I_{\rm c}}{I_{\rm c}}(\frac{\Delta\nu}{\nu})_{\rm out}.
\end{equation}
If $I_{\rm c}=I_{\varepsilon}\sim 17\%I$, typical glitch size of the Crab puslar will be $(\Delta\nu/\nu)_{\rm out}=(\Delta\nu/\nu)_{\rm typical}\sim 2\times 10^{-8}$.
The corresponding glitch originates from the inner part of the Crab pulsar will have a size
\begin{equation}
(\frac{\Delta\nu}{\nu})_{\rm in}\sim 10^{-7},
\end{equation}
consistent with the sizes of the above two relatively large glitches in the Crab pulsar.
Note that, this estimation includes only $\varepsilon$ effect, the glitch size will be larger if the over recovery is considered.

Another important consequence of this model is that, the high density layer may serve as a reservoir of oblateness in the following evolution.
Previous starquake model assumed the solid quark star has an uniform density $\overline\rho$~\citep{Zhou2004, Zhou2014, Lai2018}, thus the change rate of the oblateness of the whole star is $|\dot\varepsilon|\propto 1/\overline\rho$.
The equilibrium oblateness of the star, as shown in Eq. (\ref{ob-equilibrium}), is also calculated based on the uniform density assumption.
We assume the average density of the low and high density layer is $\overline\rho_{\rm out}$ and $\overline\rho_{\rm in}$, $\overline\rho_{\rm out}<\overline\rho<\overline\rho_{\rm in}$.
If the difference in density throughout the star is considered, the high density layer will actually decrease the oblateness a bit slower than the uniform star, i.e., $|\dot\varepsilon|_{\rm in}<|\dot\varepsilon|$.
The difference between $|\dot\varepsilon|_{\rm in}$ and $|\dot\varepsilon|$ may be small, but as it accumulates as the pulsar ages, the oblateness of high density layer will be much higher than that of the uniform star, thus the high density layer may act as a reservoir of oblateness in the Vela-like middle-aged pulsars.

\subsection{starquake in Vela-like middle-aged pulsars}
\begin{figure}
\centering\resizebox{\hsize}{!}{\includegraphics{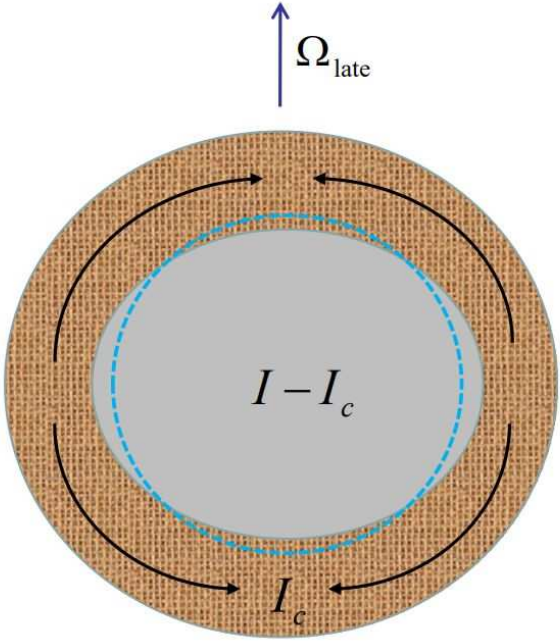}}
\caption{This figure shows the difference in oblateness between the small stones (khaki) in the outer part and the big stone (grey) in the inner part of a Vela-like middle-aged pulsar with the angular spin velocity $\Omega_{\rm late}$.
The innermost dashed ellipsoid (light blue) represents the reference oblateness at the spin frequency of the Vela pulsar.
The directions of the black solid arrows represent that, matter can move from the equatorial region to the polar region.
One difference between Fig.{\ref{fig:crab-like}} and this figure is that, we assume most of the matter can not move back towards the equatorial region.
}
\label{fig:vela-like}
\end{figure}

As pulsars spin down continually, Crab-like young pulsars become Vela-like middle-aged ones gradually.
The starquake process in Vela-like pulsars is illustrated in Fig.{\ref{fig:vela-like}}.
The physical process of starquake in the Vela-like middle-aged pulsars is the same with that in Crab-like young pulsars, but there are at least four structural differences between the Crab-like and Vela-like pulsars.
Firstly, the moment of inertia of the small stones in the outer part will be much larger than that in the Crab-like young pulsars from the view of evolutionary, which means that more matter may be involved in matter motion from the equatorial region to the polar region.
Secondly, as the spin down rate of the Vela pulsar is more than one order of magnitude lower than that of the Crab pulsar, the difference in change rate of oblateness between the high and low density layers of the Vela pulsar has decreased dramatically. Therefore, both the inner and outer parts will contribute to $\varepsilon$ effect.
Thirdly, the oblateness of the inner part is larger than that of the outer part, thus the inner part may provide more oblateness change.
Most importantly, more matter may be involved in matter motion from the equatorial region to the polar region, if the energy release during glitch is insufficient to provide its kinetic energy, part of them will not move back towards the equatorial region and pile up at the polar region.
This process redistributes matter and reduces the degree of density uniformity, resulting in a relatively large $\eta$ effect and helping to understand its typical glitch size of $10^{-6}$ and its small recovery coefficient of $Q\sim (1-5)\times 10^{-3}$.
Unfortunately, we are unable to quantitatively determine the relation between $\varepsilon$ and $\eta$ effects at present.

To summarize, typical glitch size of the Vela pulsar shall be much larger than that in the Crab pulsar from three aspects.
On the one hand, the whole star contributes to $\varepsilon$ effect.
On the other hand, the inner part can serve as a reservoir of oblateness.
Moreover, $\eta$ effect enlarges glitch size one step further.

\subsection{Energy releases during glitches}
The energy release during the glitch epoch is another interesting topic.
The energy release of a solid quark star includes the release of the gravitational energy and the strain energy.
According to Lai et al.~\citep{Lai2018},
the gravitational energy release is
\begin{equation}
\Delta E_{\rm grav}=2A\varepsilon_{+1} \Delta\varepsilon,
\end{equation}
and the strain energy release is
\begin{equation}
\Delta E_{\rm strain}=2B(\varepsilon_{0}-\varepsilon_{+1})(\Delta\varepsilon_{0}-\Delta\varepsilon)
\end{equation}
where $\varepsilon_{0}$ is the reference oblateness, $\Delta\varepsilon_{0}$ is the reference oblateness change, $\varepsilon_{+1}$ is the oblatebess of the solid quark star right before the glitch and $\Delta\varepsilon$
is the actual oblateness change of the solid star.
Besides,
\begin{equation}
\Delta\varepsilon=\frac{B}{A+B}\Delta\varepsilon_{0}.
\end{equation}
In total, the energy release is
\begin{equation}
\label{energy}
\Delta E=\Delta E_{\rm grav}+\Delta E_{\rm strain}=2A\varepsilon_{0}\Delta \varepsilon.
\end{equation}
The energy release in our model is totaly the same in formulae with those presented by Lai et al.~\citep{Lai2018}, the difference lies in
the estimated actual oblateness change $\Delta\varepsilon$.
Lai et al. estimated the actual oblateness change during glitch according to the time interval $\Delta t_{q}$ between two
successive glitches, i.e.,
\begin{equation}
\Delta t_{q}=\frac{A(A+B)}{BI_{0}}\frac{\Delta\varepsilon}{2\pi^{2}\nu|\dot\nu|}.
\end{equation}
Based on the shear modulus that $\mu\sim 10^{32}~\rm erg/cm^{3}$, i.e., $B\sim 10^{-2}A$,
Lai et al. estimated that typical actual oblateness change is $\Delta\varepsilon=10^{-10}$ for the Crab pulsar and $\Delta\varepsilon=10^{-11}$ for the Vela pulsar.

However, as we have presented in section 3, in order to reproduce the linear glitch activity of all pulsars, $B\simeq A$ is required, which means that $\mu\simeq 3\times 10^{34}~\rm erg/cm^{3}$.
In this case, the actual oblateness change should be
\begin{equation}
\Delta\varepsilon=\frac{B}{A+B}\Delta\varepsilon_{0}=\frac{B}{A+B}|\dot\varepsilon_{0}|\Delta t_{q}.
\end{equation}
Assuming $M=1.4~M_\odot$ and $R=10~\rm km$ for the Vela pulsar, according to Eq. (\ref{energy}), typical energy release during a glitch in the Vela pulsar is
\begin{equation}
\Delta E_{\rm vela}\sim 2.4\times 10^{40}(\frac{\varepsilon_{0}}{2.2\times10^{-5}})(\frac{\Delta\varepsilon}{8.8\times 10^{-9}})~\rm erg,
\end{equation}
which is three orders of magnitude larger than that estimated by Lai et al.~\citep{Lai2018}.
For the Crab pulsar, as we stated in section 3.2, its oblateness change during normal spin down is sufficient to
explain its glitch size and glitch activity if about $17\%$ of its total moment of inertia is involved in oblateness change
and stress release, this equals to an effective oblateness change of the whole star through
\begin{equation}
I\Delta\varepsilon_{\rm eff}\sim 17\% I \frac{B}{A+B}|\dot\varepsilon_{0}|\Delta t_{q-\rm crab}.
\end{equation}
Assuming $M=1.4~M_\odot$ and $R=10~\rm km$ for the Crab pulsar, we obtain
\begin{equation}
\Delta\varepsilon_{\rm eff}\sim 2.13\times 10^{-8},
\end{equation}
therefore, according to Eq. (\ref{energy}), typical energy release during a glitch in the Crab pulsar is
\begin{equation}
\label{eq55}
\Delta E_{\rm crab}\sim 4.2\times 10^{41}(\frac{\varepsilon_{0}}{1.58\times10^{-4}})(\frac{\Delta\varepsilon_{\rm eff}}{2.13\times 10^{-8}})~\rm erg.
\end{equation}
The small glitch in the Crab pulsar will release more energy than the large glitch in the Vela pulsar, which seems to be contradictory to the common sense that large glitches are accompanied by more energy release.
However, it should be pointed out that, $\eta$ effect is assumed to contribute no energy release~\citep{Lai2018}.
Therefore, $\Delta E_{\rm vela}$ and $\Delta E_{\rm crab}$ represent the lower limit of energy release for glitches in the Vela and Crab pulsars respectively.
The main reason that small glitch in the Crab pulsar will release more energy than that in the Vela pulsar is that, the reference oblateness of the Crab pulsar is nearly one order of magnitude higher.

\subsection{The recovery coefficient $Q$}
Matter motion from the equatorial region to the polar region has two consequences.
On the one hand, it results in $\eta$ effect and contributes to glitch size.
On the other hand, it affects the recovery coefficient $Q$ through two totally opposite ways.
Firstly, a small amount of matter moves which results in no significant $\eta$ effect, in this case, $Q\lesssim 1$.
Secondly, a large amount of matter moves which results in significant $\eta$ effect, however, most of the matter does not move back to the equatorial region, in this case, $Q\ll 1$.
These two cases correspond to the Crab and Vela pulsars separately.
We try to qualitatively explain $Q$ values of the Crab and Vela pulsars through the combination of $\varepsilon$ and $\eta$ effects.

For the Crab pulsar, $Q\lesssim 1$.
According to Eq.(\ref{Qdefinition}),
$Q=(\Delta\varepsilon_{\rm m}-\Delta\varepsilon)/(\Delta\varepsilon_{\rm m}+\Delta\eta)$.
Considering the average over recovery factor $k\sim 10$ and ignoring $\eta$ effect, i.e., $\Delta\eta=0$, $Q\sim 0.9$.
In our starquake model, the high recovery coefficients for glitches in the Crab pulsar indicates that no significant $\eta$ effect exists in the Crab pulsar.
This could be understood through two aspects.
Firstly, the Crab pulsar is more oblate than the Vela pulsar, its mass density around the cracking place may be a bit lower than that in the polar region.
When the star cracks in the equatorial region, most of the matter is simply compressed to eliminate the density difference between the equatorial region and the polar region.
Secondly, even if a small amount of matter moves from the equatorial region towards the polar region, the total energy release during glitch can accelerate a mass of $10^{-8}M_{\odot}$ to $c/100$ according to Eq.(\ref{eq55}), $c$ is the speed of light in vacuum.
So this part of matter may collide at the polar region and then return back to the equatorial region.

\begin{figure}
\centering
\includegraphics[width=0.50\textwidth]{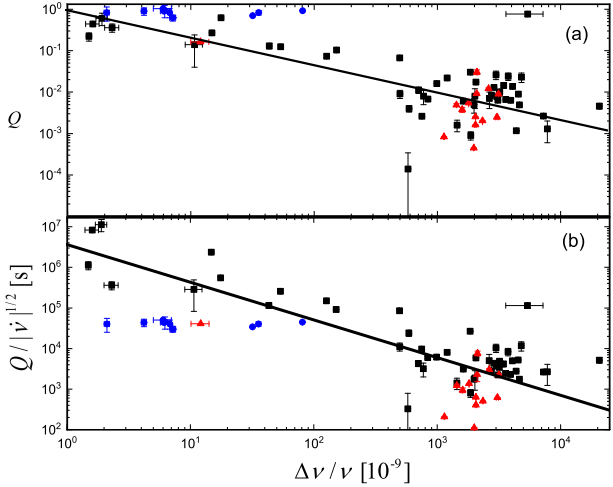}
\caption{(a)The updated recovery coefficient $Q$ versus glitch size distribution for all glitches with measured $Q$.
The blue points correspond to glitches in the Crab pulsar, the red triangles correspond to glitches in the Vela pulsar, and the black squares correspond to glitches in all other pulsars.
The black solid line shows approximately that, $Q$ decreases as the glitch size increases for all glitches except that in the Crab pulsar.
(b)
The reduced recovery coefficient $Q/|\dot{\nu}|^{1/2}$ versus glitch size distribution for all glitches with measured $Q$.}
\label{fig:Q-size2}
\end{figure}

For the Vela-like pulsars, $Q\ll 1$, indicating $\Delta\eta\gg \Delta\varepsilon_{\rm m}$.
This could be understood through its structure as shown in Fig.\ref{fig:vela-like}.
The oblateness of the outer part of the Vela pulsar has decreased much lower than that of the Crab pulsar after countless compression during starquake.
As the Vela pulsar becomes more and more spherical, the density difference between the equatorial and the polar regions becomes smaller and smaller.
When the star cracks in the equatorial region, part of the matter could be pushed towards the polar region.
As the energy release during glitch in the Vela pulsar is one order of magnitude lower than that in the Crab pulsar, the fraction of matter which returns back to the equatorial region may be smaller than that in the Crab pulsar, which will result in small recovery coefficients.

These simple and naive explanations could qualitatively account for the difference in recovery coefficient between the Crab and Vela pulsars.
From the aspect of the Crab pulsar, we expect large glitches are accompanied by relatively small $Q$ because $\eta$ effect becomes more and more important as $\varepsilon$ effect increases, this could be tested by measuring the recovery coefficient of relatively large and isolated glitch in the Crab pulsar in the future.
For all glitches with measured recovery coefficients, the statistics between $Q$ and glitch size shows that,
$Q$ decreases as glitch size increases for most of the glitches, as shown in Fig.\ref{fig:Q-size2}(a).
However, the recovery coefficients of glitches in the Crab pulsar do not follow this tendency, we can see that the blue points distribute almost parallel to the horizontal axis.
This difference could arise from the fact that, the Crab pulsar is the youngest one with glitches detected and $Q$ measured, its structure could be greatly different from other ones, just as we described in our two-layered starquake model.
Although glitches in the Crab pulsar have $Q\sim 1$, there is actually a trend that $Q$ decreases as glitch size increases.
For example, the two relatively large glitches on MJD 42447.26 ($\Delta\nu/\nu=35.7(3)\times 10^{-9}$) and MJD 50260.031 ($\Delta\nu/\nu=31.9(1)\times 10^{-9}$) have $Q=0.8(1)$ and $Q=0.680(10)$ separately, while other small glitches have $Q\sim 1$.
The problem is that $Q$ does not decreases as fast as that in other pulsars.
Future measurement of $Q$ of large glitch in the Crab pulsar may test if it follows the linear tendency as shown in Fig.\ref{fig:Q-size2}(a).

In this two-layered starquake model, the crucial difference between the Crab-like young pulsars and the Vela-like middle-aged pulsars is the spin down rate.
Therefore, the recovery coefficient has dependencies on both the spin down rate and the glitch size.
We find that, after eliminating the effect of spin down rate by replacing $Q$ with the reduced recovery coefficient $Q/|\dot{\nu}|^{1/2}$, all points are approximately uniformly distributed around the linear tendency in logarithm scale, as shown in Fig.{\ref{fig:Q-size2}}(b).
This finding may possibly support our proposal that, the key difference between the Crab and the Vela pulsars lies in the spin down rate.
Besides, the linear tendency indicates that, large and small glitches may have the same physical origin.

\section{conclusion and discussion}
We have formulated the glitch activity induced by the bulk-invariable (type I) starquakes of strangeon stars, and provided an explanation for the small glitch and low glitch activity of the Crab pulsar.
Parameters $B$ and $A$ should satisfy the relation $B\simeq A$ in order to fulfill the glitch statistics, which means that shear modulus of the strangeon star should be $\mu\simeq 3\times 10^{34}~\rm erg/cm^{3}$ for a pulsar with mass $M=1.4~M_\odot$ and radius $R=10~\rm km$ according to Eq.(\ref{eq31}).
If a pulsar with mass
$M=1.4~M_\odot$ has a largest radius $R=13.6~\rm km$~\citep{Annala2018},
the corresponding average shear modulus decreases to $\mu\simeq 10^{34}~\rm erg/cm^{3}$.
Both of them are consistent with the upper limit presented by Zhou et al.~\citep{Zhou2004} in order of magnitude.
It is worth mentioning that, the shear modulus should has a radial profile due to density gradient.
However, the density difference between the surface and the center may be a factor of $3$ for the solid strangeon star, moreover, $\mu$ is derived from the relation $B\simeq A$ where $B$ represents a global property if we assume almost the whole star is involved in the glitch process in Vela-like pulsars, so strictly speaking, the above $\mu$ represents an average or effective value.
However, even if $B\simeq A$, more oblateness change is needed. %
Our estimations show that, about $5-10$ times that accumulated during the time interval between two successive glitches is required for most of glitches in the Vela pulsar and other middle-aged pulsars respectively.

In this paper, the formulation of starquake is based on the semi-Newtonian approach, similar to that adopted by Baym \& Pines 1971.
However, as the gravitational field of the NS is pretty strong, general relativistic effects are large enough to modify the NS structure and affect the evaluation of changes in the moment of inertia, which is crucial in the starquake theory of pulsar glitch~\citep{Quintana1976}.
Particularly, if parameter $B$ is almost equal to parameter $A$ for the rigid relativistic sphere in the strangeon star model, the general-relativistic elasticity is needed to describe the solid matter in the strong gravitational field~\citep{Carter1972}.
Given all these considerations, our discussions about the shear modulus can only be treated as order-of-magnitude analysis.
Moreover, the general-relativistic effect and the relativistic nature of the elastic energy would affect the radial profile of shear.

The requirement of more oblateness change in the Vela and other middle-aged pulsars, together with the small glitch sizes and low glitch activity in the Crab pulsar, motivate the construction of the phenomenological and rough density-dependent two-layered starquake model in solid quark stars, illustrated by Figs. \ref{fig:crab-like} and \ref{fig:vela-like}.
The stars are divided into two parts, the outer part composed of small stones, and the inner part composed of a large stone.
The relatively large glitches in the Crab pulsar and the large glitches in the Vela-like pulsars can be qualitatively explained, both the typical small glitches and the low glitch activity of the Crab pulsar can be understood if the moment of inertia of the outer part reaches $\sim 17\%I$ (assuming $M=1.4~M_{\odot}$ and $R=10~\rm km$).
Besides, recovery coefficients of glitches in the Crab and Vela pulsars can also be qualitatively explained by the combination of $\varepsilon$ and $\eta$ effects.
We have to admit that, there is no way to quantitatively determine the relation between $\varepsilon$ effect and $\eta$ effect at present because of our ignorance of properties of strangeon matter.

The energy releases accompanied by glitches in the Vela and the Crab pulsars have also attracted many attention. As far as we know, the Jodrell Bank Observatory~\citep{Shaw2018}, the Neutron star Interior Composition Explorer (NICER)~\citep{Vivekanand2020}, the X-Ray Pulsar Navigation-I (XPNAV-1)~\citep{Zhang2018}, and the PolarLight onboard CubeSat~\citep{Feng2020} have searched for changes in X-ray fulx, but no changes have been identified so far.
Our estimations about the energy releases in typical glitches of the Vela and Crab pulsars are about three orders of magnitude higher than previous results~\citep{Zhou2014, Lai2018},
as we have used larger value of shear modulus derived from the glitch activity, which corresponds to the larger oblateness change during a glitch.

It is still unclear how the energy will be released, however, at least four channels could be involved in the dissipation process.
Firstly, right after the glitch, starquake may excite some oscillation modes and induce short-timescale gravitational waves (GW) or gravitational wave burst~\citep{Keer2015, Layek2020}.
Secondly, starquake may lead to energetic particle outflow or magnetic reconnection, producing radiative changes such as an fast radio burst (FRB)~\citep{Wang2018} shortly after the glitch.
Thirdly, the released energy may melt and heat the outermost layer of NS and then be converted into X-ray emission during the subsequent post glitch relaxation through cooling.
If this channel dominates the energy dissipation process, NICER, XMM-Newton and Nuclear Spectroscopic Telescope Array (NuSTAR) may have a chance to detect this X-ray enhancement in some X-ray faint rotation powered pulsars (i.e., weak X-ray background) at the very beginning of post glitch relaxation.
Finally, starquake may excite some kinds of asymmetry in NS structure and form the so-called mountains, and the unreleased elastic energy may be dissipated through transient gravitational waves during post glitch relaxation~\citep{Yim2020, Gao2020}.
Anyway, further investigation on the detailed energy release mechanisms is worth performing.

How to distinguish the superfluid model in conventional NS from the starquake model in the solid strangeon star?
Among the several aspects in the context of glitch, i.e., glitch rise, glitch recovery, radiative and pulse profile changes, and GW emission, the latter two are most promising.
(1) Ashton et al. has set an upper limit of $12.6~\rm{s}$ on the glitch rise timescale for the 2016 glitch in the Vela pulsar~\citep{Ashton2019}, however, a starquake may occur in a timescale of several milliseconds, thus this upper limit is not tight enough to serve as a criterion.
(2) The glitch recovery is complicated and strongly model dependent (for example, multi components with different pinning energy is needed to fit the recovery process in the superfluid model), it can hardly serve as a clear probe.
(3)Vortex creep in the post glitch recovery process in the superfluid model will results in energy release due to the friction between the superfluid and normal components~\citep{Alpar1984}, but this energy should be emitted thermally, besides, there should be a time delay between detection of the glitch and soft X-ray enhancement (if detected) because of the relatively low thermal conductivity in NS crust.
In the strangeon star, matter movement in the surface may affect the magnetic field lines, resulting in radiative and pulse profile changes instantly after the glitch, pulse profile change is supposed to be accompanied with every glitch.
If the association of the glitch with short-timescale radiative change such as the FRB be confirmed in the future, it will strongly support the origin of glitch as a starquake.
Besides, there is no need for the radiative change to occur in the soft X-ray band, moreover, it can be emitted non-thermally.
(4)Both the conventional NS and the strangeon star could generate GW burst instantly after the glitch~(timescale $\sim\rm{ms}$)~\citep{Keer2015} and/or continuous GW emission in the kilohertz (kHz) band~\citep{Gao2020}, however, for a specific pulsar with well-measured distance, the strain amplitude will be different due to their difference in shear modulus between these two kinds of stars.
The conventional NS could also generate GW emission through \emph{f}-mode oscillations in the $\rm{kHz}$ band~(timescale $\sim\rm{100~ms}$)~\citep{Ho2020}.
On the other hand, the solid strangeon star cloud generate transient GW emission through mountains at two times the spin frequency in a timescale comparable to the post glitch relaxation timescale~\citep{Yim2020}.
To sum up, the short-timescale GW burst, continuous and transient GW could all serve as the criterion from the view of the strain amplitude, timescale and frequency of the GW emission associated with the glitch.

The forthcoming Square Kilometer Array (SKA) may be the most promising tool to study glitch in the near future due to its exquisite timing precision and high observational cadence from at least three aspects.
Firstly, its capability to improve both the number of pulsars monitored and the cadence can greatly enlarge the glitch samples.
Secondly, the detection of small glitches and the lower end of the glitch size can shed light on the glitch mechanism~\citep{Watts2015}.
And thirdly, its superior sensitivity may possibly uncover if all glitches are accompanied by radiative and/or pulse profile changes shortly after the glitch.
Other X-ray telescopes can act as the supplementary by tracking the X-ray flux change during the following recovery process.
Quick response, continuous and multi-messenger observations soon after the glitch should be vital to uncover secrets behind glitch eventually.

\section*{Acknowledgements}
We appreciate the anonymous referee for specific comments and suggestions, which have improved this manuscript.
We are also grateful to Ming-yu Ge from Institute of High Energy Physics, Chinese Academy of Sciences and Liang-duan Liu from Peking Normal University, for their useful discussions.
This work is supported by the National Key R\&D Program of China (Grant No. 2017YFA0402602), the National Natural Science Foundation of China (Grant No. 11673002, U1531243, 11773011 and U1831104), and the Strategic Priority Research Program of Chinese Academy Sciences (Grant No. XDB23010200).
XL is also supported by the Science and technology research project of Hubei Provincial Department of Education (No. D20183002).

\section*{ORCID iDs}
Lai, X. Y. https://orcid.org/0000-0002-3093-8476
\\
Wang, W. H. https://orcid.org/0000-0003-1473-5713
\\
Xu, R. X. https://orcid.org/0000-0002-9042-3044
\\
Zhou, E. P. https://orcid.org/0000-0002-9624-3749

\section*{Data availability}
The data underlying this article are available in this article.

\end{document}